\documentclass[aip,reprint]{revtex4-1}

\usepackage{graphicx}
\usepackage{dcolumn}
\usepackage{bm}
\usepackage{xcolor}
\usepackage{here}
\usepackage{multirow}
\usepackage{amsthm}
\usepackage{amsfonts}
\usepackage{amsmath}
\usepackage{hyperref} 
\usepackage{pdfpages}
\usepackage{pgffor}
\usepackage{ulem}

\makeatletter
\AtBeginDocument{\let\LS@rot\@undefined}
\makeatother

\draft 

\begin{document}


\title{Chaotic stochastic resonance in Mackey--Glass equations} 

\author{Eiki Kojima}
\email{kojima.eiki.h9@elms.hokudai.ac.jp }
\affiliation{Department of Mathematics, Hokkaido University, N12 W7 Kita-ku, Sapporo, 0600812 Hokkaido, Japan}
\author{Yuzuru Sato}
\email{ysato@math.sci.hokudai.ac.jp}
\affiliation{RIES-MSC / Department of Mathematics, Hokkaido University, N12 W7 Kita-ku, Sapporo, 0600812 Hokkaido, Japan}
\affiliation{London Mathematical Laboratory, 14 Buckingham Street, London WC2N 6DF, United Kingdom}

\date{\today}

\begin{abstract}
Stochastic resonance (SR) manifests as switching dynamics between two quasi-stationary states in the stochastic Mackey--Glass equation. We identify chaotic SR, arising from the coexistence of resonance and chaos in stochastic dynamics. In contrast to classical SR, which is described by a random point attractor with a negative largest Lyapunov exponent, chaotic SR is described by a random strange attractor with a positive largest Lyapunov exponent. We observe chaotic SR in the Mackey-Glass equation as well as chaotic SR in the Duffing equation and the underdamped FitzHugh-Nagumo equation, demonstrating the universality of this phenomenon across a broad class of strongly nonlinear random dynamical systems.
\end{abstract}


\maketitle 

\begin{quotation}
Stochastic resonance (SR) is a noise-induced phenomenon in which the response of a nonlinear system to an external forcing is enhanced at an optimal noise intensity. 
In the present work, we collectively
refer to these phenomena as “SR,” in the broad sense of
resonance phenomena arising in stochastic dynamics. 
SR can also arise in time-delayed systems, where the delayed feedback plays the role of an external forcing. In this study, we investigate the stochastic Mackey--Glass equation, a chaotic time-delayed dynamics subject to noise, and identify two distinct types of SR, namely \textit{stable SR} and \textit{chaotic SR}. To clarify these properties, 
we employ concepts in random dynamical systems theory, which enables us to directly analyze the intrinsic dynamical structure in stochastic dynamics. Stable SR is described by random point attractors with a negative largest random LE, whereas chaotic SR is described by random strange attractors with a positive largest random LE. We observe chaotic SR in the Mackey-Glass equation as well as chaotic SR in the Duffing equation and  the underdamped FitzHugh-Nagumo equation, demonstrating the universality of this phenomenon across a broad class of strongly nonlinear random dynamical systems.
\end{quotation}

\section{Introduction}
Adding noise to a deterministic dynamical system can induce qualitative changes in its behavior, known collectively as noise-induced phenomena. Stochastic resonance (SR) \cite{SR_review} is a representative example, in which the response of a nonlinear system to an external forcing is enhanced at an optimal noise intensity. SR was originally investigated in bistable systems subjected to noise and periodic forcing\cite{SR_theory, SR_climate}, where the addition of noise enhances the periodicity of the system's dynamics. Since then, SR has been observed in a wide variety of nonlinear systems. Even in the absence of periodic forcing, noise-induced periodicity can emerge, a phenomenon known as coherence resonance \cite{Pikovsky_Kurths}. The interplay between time delay and noise can also generate a coherence resonance in time-delayed systems \cite{Ohira_Sato,Tsimring_Pikovsky,Masoller}. In the present work, we collectively refer to these phenomena as “SR,” in the broad sense of resonance phenomena arising in stochastic dynamics.

SR has also been reported in chaotic systems 
\cite{carroll1993stochastic, nicolis1993stochastic,crisanti1994stochastic, reibold1997stochastic,jungling2008noise,anishchenko1993stochastic, anishchenko1992stochastic,zakharova2010stochastic}. Most previous studies have been conducted within the framework of deterministic dynamical systems theory, since the stochastic process approaches\cite{SR_review, SR_theory, mcnamara1989theory, wiesenfeld1994stochastic, gingl1995non} does not focus on the  characterization of the underlying dynamical structure of the system. 
Deterministic dynamical systems approaches primarily address resonance phenomena in the absence of external noise by interpreting deterministic chaos as an effective noise source\cite{carroll1993stochastic, nicolis1993stochastic,crisanti1994stochastic, reibold1997stochastic,jungling2008noise}. These studies have shown that SR can be described in terms of crisis. Furthermore, extensions of stochastic approaches to deterministic crises and noise-induced crises\cite{sommerer1991scaling} have demonstrated that chaotic systems can exhibit stochastic multiresonance\cite{vilar1997stochastic}, in which the response to periodic forcing is maximized at multiple distinct noise intensities \cite{krawiecki2001noisefree, matyjaskiewicz2003stochastic}.

In this paper, we employ random dynamical systems theory, which enables a direct analysis of the intrinsic dynamical structure in stochastic systems, including random attractors, stochastic bifurcations, and stability measured by random Lyapunov exponents (LEs). Within this framework, we identify {\it stable SR}, described by random point attractors with a negative largest random LE, and {\it chaotic SR}, described by random strange attractors with a positive largest random LE, in Mackey--Glass equations. 

This paper is organized as follows. Section~\ref{sec:SMG} introduces the stochastic Mackey--Glass equation and its phenomenology. Section~\ref{sec:Chaotic_SR} presents numerical and analytical results on stable and chaotic SR. 
Section
~\ref{sec:Conclusion} provides concluding remarks.

\section{Model and phenomenology}
\label{sec:SMG}
\subsection{Stochastic Mackey--Glass equations}
Delay differential equations (DDEs) are widely used to model systems where time delay plays a crucial role \cite{Ikeda_Daido_Akimoto,  Ikeda_Matsumoto,Suarez_1988, Keane_2017,MG, Longtin_1988,Milton_1989, roxin2005role}. Despite their deceptively simple formulations, DDEs are infinite-dimensional dynamical systems capable of exhibiting rich behaviors, including high-dimensional chaos. The Mackey--Glass equation (MG) is a well-known DDE, developed as a model of hematopoiesis \cite{MG}, and is given by
\begin{equation}
    \frac{dx(t)}{dt}=\frac{ax(t-\tau)}{1+x^{c}(t-\tau)}-bx(t),
    \label{eq:MG}
\end{equation}
where $\tau >0$ denotes the delay time, $a>0$ is the strength of delayed feedback, $b>0$ is the decay rate, and $c$ is the shape parameter of the delayed feedback. The initial condition is defined as $x(t) = \phi(t)$ for $t \in [-\tau,0)$ and is assumed to be constant unless otherwise specified.
The MG can exhibit multiple positive LEs, and the effective dimension of the dynamics increases as $\tau$ grows \cite{Farmer1982}. 

To investigate the impact of external noise, we introduce the stochastic Mackey--Glass equation (SMG) by incorporating additive Gaussian noise into Eq.~\eqref{eq:MG}:  
\begin{equation}
    \label{eq:SMG}
    dx(t) = \Big[ \frac{ax(t-\tau)}{1+x^{c}(t-\tau)} - bx(t) \Big] dt + \sigma dW_t,
\end{equation}
where $\sigma$ represents the noise intensity and $W_t$ denotes a Wiener process. The parameters are fixed at the standard values $b=0.1$ and $c=10$, while $a$ and $\sigma$ serve as control parameters. The delay time is chosen to be sufficiently larger than the response time  \cite{Mensour_Longtin} $1/b$ and is set to $\tau=90$. Numerical simulations are performed using the Euler--Maruyama method with a time step of $\Delta t = 0.01$, unless otherwise stated.

\subsection{Bifurcation in the stochastic Mackey--Glass equation}
\begin{figure}[hbpt]
    \centering
    \includegraphics[scale=0.5]{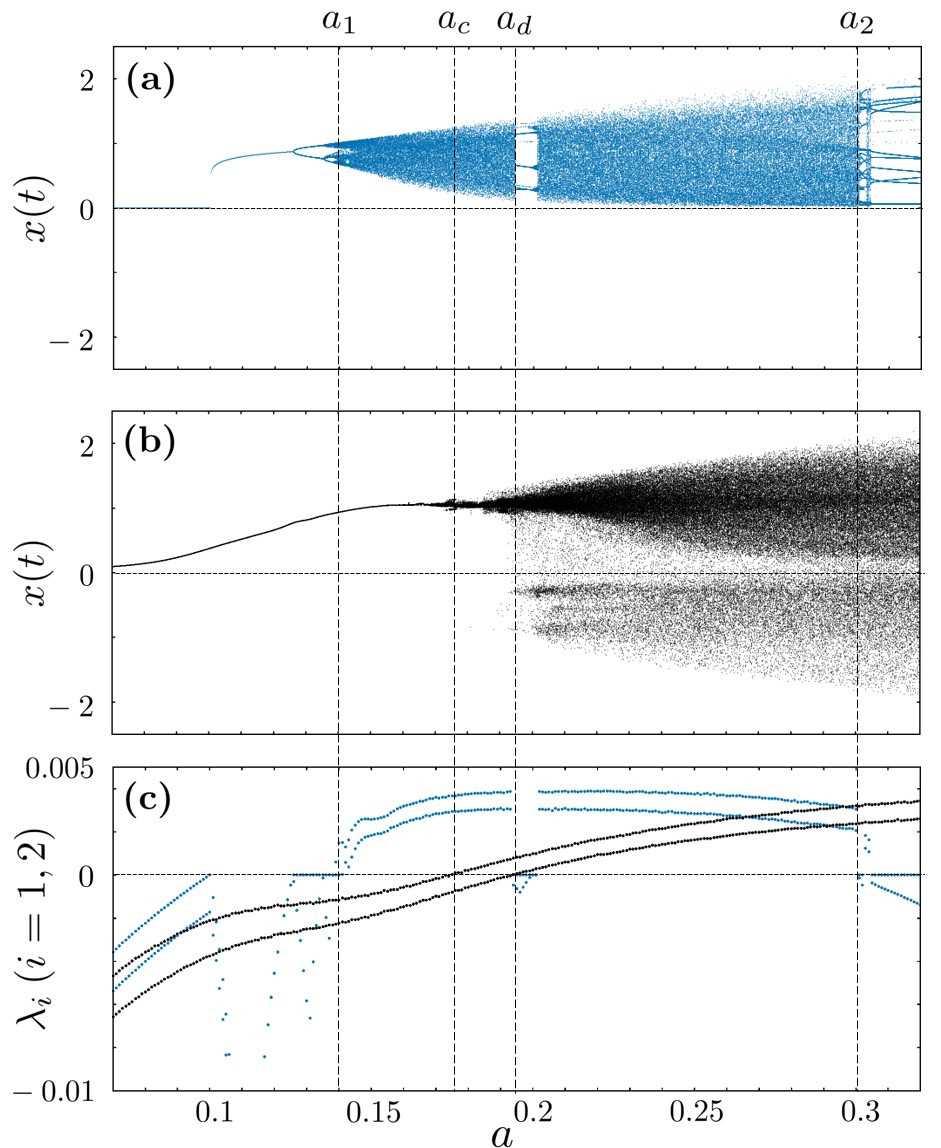}
    \caption{
    (a) Bifurcation diagram of the MG  with $\sigma = 0$. (b)  Bifurcation diagram of random pullback attractors of the SMG with $\sigma = 0.15$. (c) First and second largest LEs for $\sigma = 0$ (blue points) and $\sigma = 0.15$ (black points).
    In the bifurcation diagram, random pullback attractors are computed using $1.0\times 10^2$ constant initial functions uniformly distributed in $[-1,1]$ with a pullback time $t_p = 4\times10^4$. We designate $a_1=0.14$ as a representative example of a weakly nonlinear regime and $a_2=0.30$ as that of a strongly nonlinear regime. At $a=a_1, a_2$, low- and high-dimensional deterministic chaos are observed, respectively. The values, $a_c$ and $a_d$, indicate the zero-crossing points of the first and second largest LEs in the presence of noise. The LEs and the bifurcation diagram were computed using a time step of $\Delta t=0.1$ and a resolution defined as $N := \tau/\Delta t = 900$. Note that for delay differential equations, the largest LEs are known to be well approximated with a limited resolution of $N= \tau/\Delta t\sim 100$\cite{wernecke2019chaos}.}
    \label{fig:Bifur}
\end{figure}

We briefly summarize the phenomenology of the dynamics in the MG and SMG. In the bifurcation diagram of the MG (Fig. \ref{fig:Bifur} (a)), the Poincaré section, $dx/dt=0$, of the attractor in Eq.~\eqref{eq:MG} is plotted as a function of the delayed feedback strength $a$ (blue dots). In the bifurcation diagram of the SMG (Fig. \ref{fig:Bifur} (b)), projections of the random pullback attractors 
are shown as black dots (see Appendices~A and~B for details on pullback attractors and bifurcation diagram). The first and second LEs of the MG and SMG as functions of $a$ at $\sigma = 0.0$ and $\sigma = 0.15$ are shown in Fig. \ref{fig:Bifur} (c) (blue and black, respectively). For the MG, the origin is stable for $0<a<0.1$.  A pitchfork bifurcation occurs at $a=0.1$, giving rise to two symmetric fixed points with respect to the origin.  As $a$ increases, a Hopf bifurcation emerges at $a\simeq 0.125$, producing a limit cycle. A bifurcation to chaos follows at $a\simeq 0.138$, with high-dimensional chaos appearing around $a\simeq 0.144$. In the SMG, the zero-crossing point of the largest LE occurs at $a=a_c\simeq 0.175$, beyond which 
stochastic chaos\cite{Chekroun_Ghil} is observed. At $a=a_d\simeq 0.195$, the second largest LE becomes positive, indicating the emergence of  high-dimensional stochastic chaos. 

In summary, although the MG becomes multi-stable via a pitchfork bifurcation, the SMG exhibits a single attractor with two quasi-stable states  as two deterministic attractors are connected by external noise. As $a$ increases, the random LEs of the SMG grow monotonically, leading to stochastic chaos and later to high-dimensional stochastic chaos. Notably, these stochastic bifurcations occur at significantly larger $a$ values than in the deterministic system.

\subsection{Stochastic resonance in the stochastic Mackey--Glass equation}
\begin{figure}[htbp]
    \centering
    \includegraphics[width=0.95\linewidth]{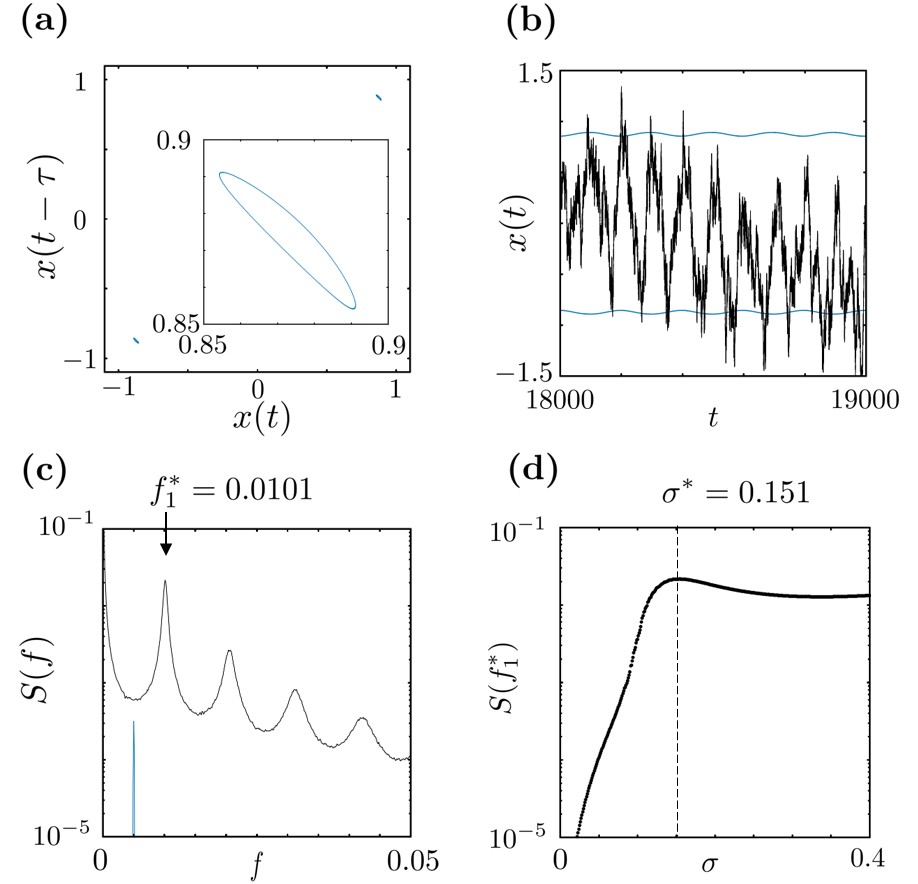}
    \caption{(a) Two deterministic attractors projected onto $(x(t),x(t-\tau))$ plane for $a=0.126$. The inset shows a magnified view of the limit cycle. (b) Time series for the deterministic case, $\sigma=0$ (blue dots), and the resonance case, $\sigma = \sigma^*$ (black dots). (c) Power spectra for $\sigma=0$ (blue lines) and the optimal noise intensity $\sigma=\sigma^*$ (black lines). (d) Power at the primary resonant frequency $f_1^*$ as a function of $\sigma$.}
    \label{fig:SR}
\end{figure}

Beyond the deterministic pitchfork bifurcation point ($a>0.1$), SR appears as switching dynamics between two quasi-stable states, associated with point, periodic, and strange attractors in the MG. Figure \ref{fig:SR} illustrates SR for a regime with deterministic periodic attractors. Without noise, two symmetric limit cycles exist with respect to the origin (Fig. \ref{fig:SR} (a)), each with a period of approximately $2\tau$. In the presence of noise, switching between these limit cycles emerges, and periodicity is enhanced at an optimal noise intensity. Figure \ref{fig:SR} (c) shows the power spectra with and without noise (black and blue lines, respectively). Under optimal noise intensity, the peaks at $f=f_n^* \simeq n/\tau$ $(n=1,2,\ldots)$ reveal clear harmonics. The power at the primary resonant frequency $f_1^*$ reaches a maximum at the optimal noise intensity $\sigma=\sigma^*$, confirming the presence of SR. We refer to $\sigma^*$ as the resonance point (Fig. \ref{fig:SR} (d)). For $\tau$ sufficiently larger than $1/b$, SR is observed over a wide range of parameters, and the primary resonant frequency is approximately given by $1/\tau$. 

\section{Chaotic stochastic resonance}
\label{sec:Chaotic_SR}
\subsection{Stable and chaotic SR}
\begin{figure}[htbp]
    \centering
    \includegraphics[scale =0.65]{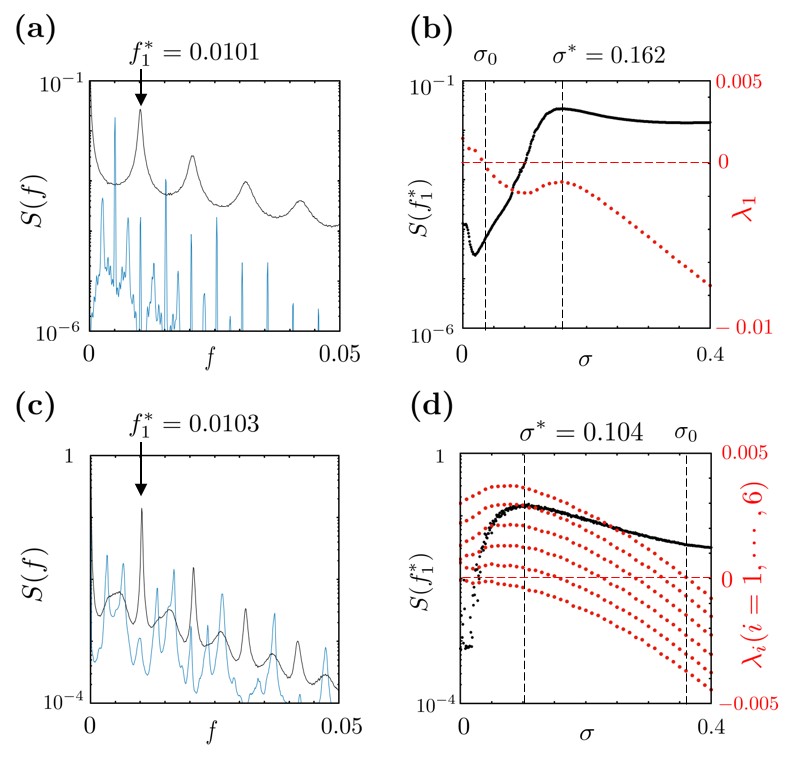}
    \caption{Power spectra for $\sigma=0$ (blue lines) and $\sigma=\sigma^*$ (black lines) at (a) $a=a_1$ and (c) $a=a_2$. Power at the primary resonant frequency $f_1^*$ (black points) as a function of $\sigma$ at (b) $a=a_1$ and (d) $a=a_2$. The largest random LE at $a=a_1$ and the largest six random LEs at $a=a_2$ are also shown in (b) and (d), respectively (red points). The random LEs were computed using a time step of $\Delta t=0.1$ and a resolution defined as $N := \tau/\Delta t = 900$.}
    
    \label{fig:Spectrum}
\end{figure}

To characterize the dynamical properties of SR, we compute the random LEs of the SMG (see Appendix~A for the definition of random LE). We examine two representative cases: one where the deterministic attractors exhibit low-dimensional chaos ($a = a_1 = 0.14$),  and one where they exhibit high-dimensional chaos ($a = a_2 = 0.30$). In both cases, switching dynamics between two quasi-stable states appears, and the power spectrum peaks at $ f^*_n \simeq n/\tau$ $(n=1,2,\ldots)$ are enhanced at the resonance point (Fig. \ref{fig:Spectrum}). The sign of the largest random LE at the resonance point distinguishes the two types of SR: a negative exponent corresponds to \textit{stable SR} (Fig. \ref{fig:Spectrum} (b)), whereas a positive exponent corresponds to \textit{chaotic SR} (Fig. \ref{fig:Spectrum} (d)).

In the case of stable SR, increasing the noise intensity causes $\lambda_1$ to become negative at $\sigma = \sigma_0$, at which point the chaotic attractor collapses into a random point attractor. As $\sigma$ is increased further, $\lambda_1$ attains a local maximum near the resonance point. Because the zero-crossing point $\sigma_0$ precedes the resonance point $\sigma^*$, chaos vanishes at resonance, rendering the system effectively analogous to the classical SR model with a double-well potential and periodic forcing \cite{SR_theory}. For sufficiently large $\sigma$, $\lambda_1$ decreases again, as trajectories stay almost always in contracting regions outside the attractors. 
In the case of chaotic SR, $\lambda_1$ initially increases and peaks near the resonance point. Here, $\sigma^*$ precedes $\sigma_0$, indicating that SR occurs while the system still possesses a positive random LE and thus retains stochastic chaos.  Chaotic SR can be regarded as a class of noise-induced phenomena that enhances the characteristic periods of stochastic chaos. The ordering of $\sigma^*$ and $\sigma_0$ provide a clear diagnostic: $\sigma_*>\sigma_0$ for stable SR and $\sigma_*<\sigma_0$ for chaotic SR.   

Stable SR, represented by  $a=a_1$, is robustly observed 
in a weakly nonlinear regime $0.1 < a < 0.15$. Chaotic SR, represented by $a=a_2$, appears in a strongly nonlinear regime $a > 0.28$, which includes deterministic window regions where noise-induced chaos \cite{mayer1981influence} arises. 
The occurrence of chaotic SR in such regimes suggests that high-dimensionality may contribute to its emergence, warranting further investigation. 

\subsection{Visualization of stable and chaotic SR}
Previous studies on SR in chaotic systems have generally not distinguished chaotic SR from stable SR \cite{anishchenko1993stochastic, anishchenko1992stochastic,carroll1993stochastic, nicolis1993stochastic,crisanti1994stochastic, reibold1997stochastic,krawiecki2001noisefree,jungling2008noise,matyjaskiewicz2003stochastic}. In this work, we characterize  stochastic dynamical structures, including stochastic chaos, within the framework of random dynamical systems. This distinction becomes evident when the dynamics are visualized using space--time representations \cite{Arrechi_1992, Yanchuk_2017} and \textit{random pullback attractors} (see Appendix~A for the construction of random pullback attractors). 

\begin{figure}[htbp]
    \centering
    \includegraphics[scale=0.48]{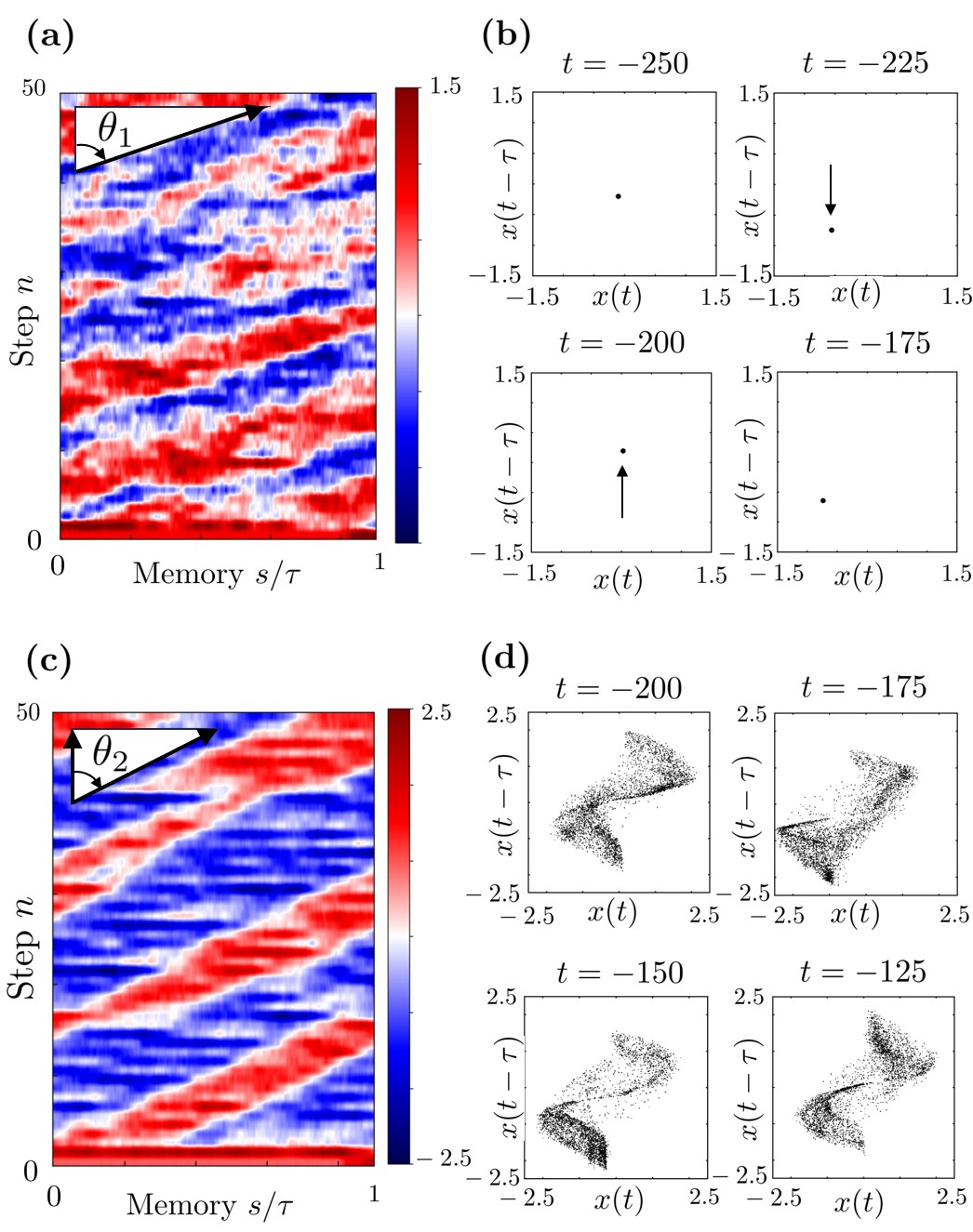}
    \caption{Space--time representation at the resonance point for (a) stable SR ($a = a_1, \sigma = 0.162$) and (c) chaotic SR ($a = a_2, \sigma = 0.104$). In both cases, traveling waves appear in the memory space $s\in[0,\tau)$. The angle $\theta$ denotes the propagation direction of the traveling wave relative to the orthogonal direction, defined by $\tan\theta=\epsilon$, where $\epsilon$ represents the deviation of the resonant period (see Eq. (\ref{eq:frequency deviation})). From the measured angles $\theta_1$ (stable SR) and $\theta_2$ (chaotic SR), we obtain $\theta_2/\theta_1 \simeq 0.786$, indicating that wave propagation is slower in the chaotic SR regime. Panels (b) and (d) show snapshots of the random pullback attractor at the resonance point computed from $1.0\times 10^5$ constant initial functions uniformly distributed on $[-1,1]$ with  a pullback time $t_p = 2\times10^4$, projected onto $(x(t), x(t-\tau))$ plane.} 
    \label{fig:Chaotic_SR}
\end{figure}

A space--time representation maps the time series onto a two-dimensional plane by introducing a discrete step $n \in \mathbb{Z}$ and a memory space $s\in[0,\tau)$. Both stable SR and chaotic SR appear as traveling waves in the memory space $s$ (Fig.~\ref{fig:Chaotic_SR}, left column). The angle $\theta$, defined as the propagation direction relative to the orthogonal direction, decreases with increasing delayed feedback strength $a$, and approaches zero as $\tau$ increases. Consequently, the angle  $\theta_1$ for stable SR is larger than $\theta_2$ for chaotic SR. Notably, chaotic SR exhibits pronounced periodicity, with chaotic internal structures within the traveling waves, indicating the coexistence of resonance and chaos in stochastic dynamics. 

When the largest random LE is negative, the random pullback attractor collapses to a random point attractor since initially close trajectories converge and eventually synchronize. 
Denoting the  distribution of all possible trajectories at time $t$ by  $\rho_t$, stable SR can be interpreted as pseudo-periodic motion of $\rho_t$ on the random point attractor (Fig.~\ref{fig:Chaotic_SR} (b)). Conversely, when the random LEs are positive, the random pullback attractor becomes a random strange attractor. In this case, chaotic SR corresponds to pseudo-periodic motion of $\rho_t$ on a random strange attractor (Fig.~\ref{fig:Chaotic_SR} (d)). Such pseudo-periodic motion of $\rho_t$ is related to the problem of statistical periodicity \cite{lasota1987noise, Losson_1995}. The precise relationship between chaotic SR and statistical periodicity remains an open question and constitutes an interesting direction for future study.

\subsection{Resonant period and unstable spiral}
\begin{figure}[htbp]
    \centering
    \includegraphics[scale =0.5]{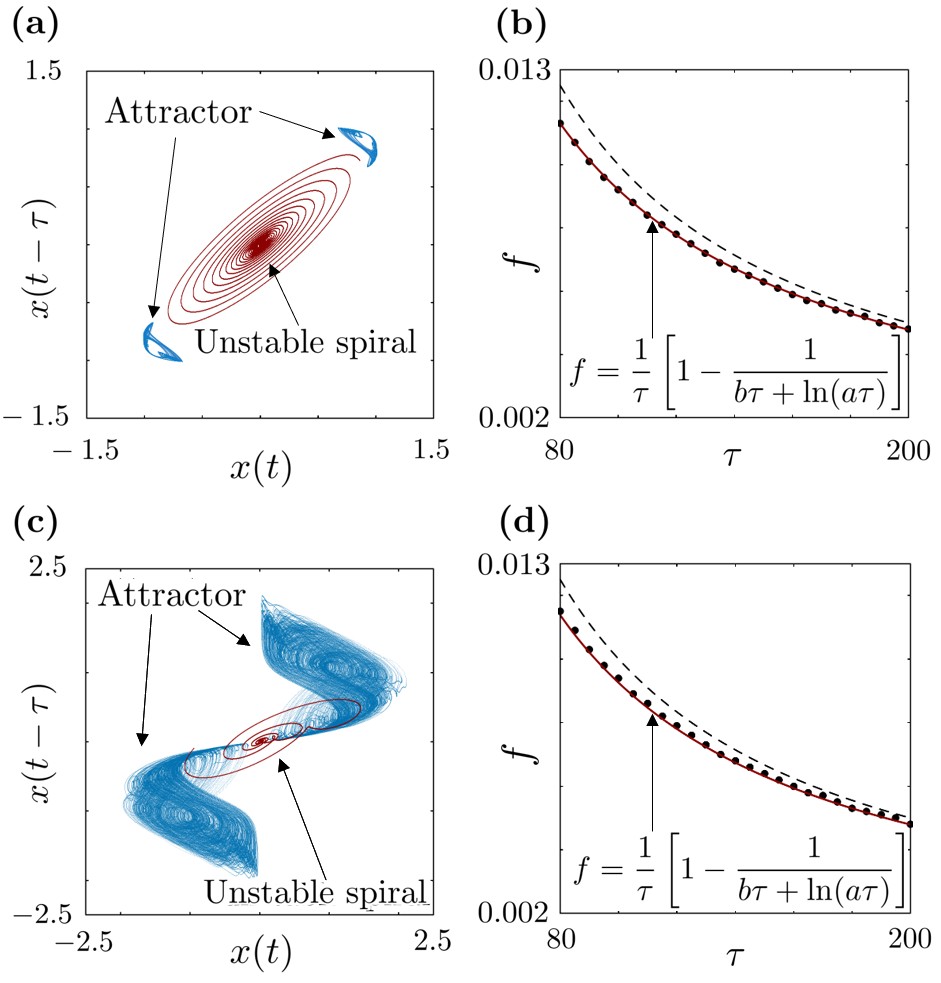}
    \caption{Unstable spirals in the weakly and strongly nonlinear regimes. Phase-space portraits projected onto $(x(t), x(t-\tau))$ plane for (a) $a = a_1$ and (c) $a = a_2$. Transients initialized near the origin (red dots) eventually converge to other stable attractors (blue dots). The trajectory is computed from the initial function $\phi(t)= 0.01 \sin(2\pi f^*_1 t) + 0.02 \sin(2\pi f^*_2 t)$ and departs from the origin following a spiral structure. The primary resonant frequency $f_1^*$ as a function of $\tau$ for (b) $a = a_1$ and (d) $a = a_2$. The theoretical estimate, the reciprocal of Eq. (\ref{eq:frequency deviation}), (red lines) agrees well with numerical results (black points). The classical estimate $f_1^* = 1/\tau$ is also shown (dotted lines).}
    \label{fig:Spiral}
\end{figure}

In studies of SR in DDEs, the resonant period $T = 1/f^*_1$ is often approximated as $T \simeq \tau$. However, the traveling waves observed in the space–-time representation (Fig. \ref{fig:Chaotic_SR}, left column) indicate that the resonance period is slightly longer than $\tau$. We refined estimate, $T = \tau (1+\epsilon)$, based on the linear mode of the unstable spiral around the origin (Fig. \ref{fig:Spiral}, left column). This approximation improves upon the classical one in the regime where the delay $\tau$ is sufficiently larger than the response time $1/b$, but not excessively large. For sufficiently large $\tau$, the frequencies of the linear modes at $x=0$ are given by \cite{Mensour_Longtin, Amann2007}
\begin{equation}
    s_{2m} = \frac{m}{\tau} \left[ 1-\frac{1}{b \tau + \ln(a\tau)} \right]
    ~~(m=0, \pm1, \cdots). 
\end{equation}
We find that the $n$th resonant frequency satisfies $f^*_n\simeq s_{2n} ~(n=1,2, \ldots)$, and thus the resonance period can be estimated as (see Appendix~C)
\begin{align}
    T = \frac{1}{f_1^*}
    \simeq \frac{1}{s_{2}} = \tau (1+ \varepsilon), ~~~~ \varepsilon = \frac{1}{b\tau + \ln(a\tau)-1}. 
    \label{eq:frequency deviation}
\end{align}
The classical approximation, $T\simeq \tau$, holds only when $\tau$ is extremely large. Numerical validation confirms that the primary resonant frequency, $f_1^*$ (black points), agrees well with the theoretical estimate, $s_2$ (red line), as shown in Figs.~\ref{fig:Spiral} (b) and (d). This result suggests that SR is associated with frequent visits to the unstable spiral caused by noise. Since our approximation is based on these spiral structures, it is applicable to a broad class of DDEs beyond the specific case of the Mackey–Glass equation. 

\section{Conclusion}
\label{sec:Conclusion}
We investigated the dynamics of the Mackey--Glass equation in the presence of noise. In the weakly nonlinear regime, stable SR emerges as switching dynamics between two quasi-stationary states, with the largest LE remaining negative at the resonance point. In this regime, the resonance point $\sigma^*$ follows the zero-crossing point of the largest LE $\sigma_0$ ($\sigma_0 < \sigma^*$). The deterministic chaotic attractor becomes stabilized into a random point attractor, and a resonance phenomenon analogous to that in the classical double-well potential model \cite{SR_theory} effectively occurs. In the strongly nonlinear regime, we identify chaotic SR, characterized by the coexistence of SR and stochastic chaos with positive LEs. Here, the resonance point $\sigma^*$ precedes the zero-crossing point $\sigma_0$ ($\sigma^* < \sigma_0$), indicating that resonance occurs while the system remains chaotic. 

We observe chaotic SR in the Mackey-Glass equation as well as 
chaotic SR in the Duffing equation and the underdamped FitzHugh-Nagumo equation (see Appendix-D), demonstrating the universality of chaotic SR across a broad class of strongly nonlinear random dynamical systems. 

\begin{acknowledgments}
Authors thank Prof. T. Ohira (Nagoya University), Prof. Hiroya Nakao (Institute of Science Tokyo), and Prof. Hiroshi Kori (University of Tokyo) for valuable comments. Authors are supported by JSPS Moonshot project No. JPMJMS2282-15.  
Y.S. is supported by JSPS Grant-in-Aid for Scientific Research (B), JP No. 21H01002.
\end{acknowledgments}

\section*{Data Availability Statement}
The data that support the findings of this study are available from the corresponding author upon reasonable request.

\appendix

\section{Definitions of random pullback attractor and random Lyapunov exponent}
To parametrize a noise realization $\omega\in \Omega$ in time $t$, we introduce a family of measure-preserving maps $\theta_t:\varOmega\rightarrow\varOmega$ satisfied with $\theta_0 = \rm{id}_{\Omega}$ and $\theta_{s+t}=\theta_s \circ \theta_{t}$ for all $t$. The time evolution of random dynamical systems is described by a stochastic flow $\varPhi(t,\omega): X\rightarrow X$ satisfying the cocycle property $\varPhi(t+s,\omega) = \varPhi(t,\theta_s(\omega)) \circ \varPhi(s,\omega)$. For systems with a single attractor, a set $\mathcal{A}(\omega)$ is defined as a random pullback attractor if it satisfies the following three conditions (see \cite{arnold1998random,Chekroun_Ghil} for the exact definition):
\renewcommand{\theenumi}{\roman{enumi}}
\begin{enumerate}
    \item 
    $\mathcal{A}(\omega)$ is compact, i.e., $A(\omega):=\{x\in X|  ~(\omega,x)\in\mathcal{A}(\omega)\} \subset X$ is compact for almost all $\omega\in\Omega$.
    \item $\mathcal{A(\omega)}$ is $\varPhi$-invariant, i.e., for all $t$, $\varPhi(t,\omega)A(\omega) = A(\theta_t\omega)$ for almost all $\omega\in\Omega$.
    \item $\mathcal{A(\omega)}$ is attracting in the pullback sense, 
    i.e., $\lim_{t\rightarrow\infty} d_X (\varPhi(t,\theta_{-t}\omega)B, A(\omega)) = 0$ holds for all $ B \subset X$ and for almost all $\omega \in \Omega$.
\end{enumerate}
where $d_X$ denotes the Hausdorff semi-distance.

\vspace{2mm}
A realization $A(\omega)$ of a random pullback attractor ${\mathcal A}(\omega)$ can be numerically approximated by a snapshot of trajectories evolving from a set of initial values $B$, under a fixed noise realization $\omega$, for a integration period given by the pullback time $t_p$. For a precise numerical computation, a large number of initial points for $B$ and a sufficiently long pullback time $t_p$ are adopted. Lyapunov exponents (LEs) for stochastic dynamics on random pullback attractors are defined as the average expansion rates of infinitesimal perturbations, in direct analogy with deterministic systems. As a simple example, the LE of a one-dimensional stochastic differential equation $dx = f(x)dt + \sigma dW_t$, where $W_t$ denotes a Wiener process, is given by
\begin{align}
    \lambda(\omega, x_0) = \lim_{T\rightarrow\infty} \frac{1}{T} \int_0^{T} f'(\varPhi(t,\omega)x_0) dt, 
\end{align}
where $\varPhi$ denotes the stochastic flow and $x_0\in X$ is the initial state. In general, $\lambda(\omega, x_0)$ is a random variable. For ergodic systems with a single attractor, $\lambda(\omega,x_0)$ takes constant value for almost all $\omega$ and for any $x_0$.
For high-dimensional random dynamical systems, the Lyapunov spectrum is defined in a manner analogous to that of high-dimensional deterministic dynamical systems.

\section{Bifurcation diagram of random pullback attractors}
\begin{figure}[htbp]
    \centering
    \includegraphics[width=0.9\linewidth]{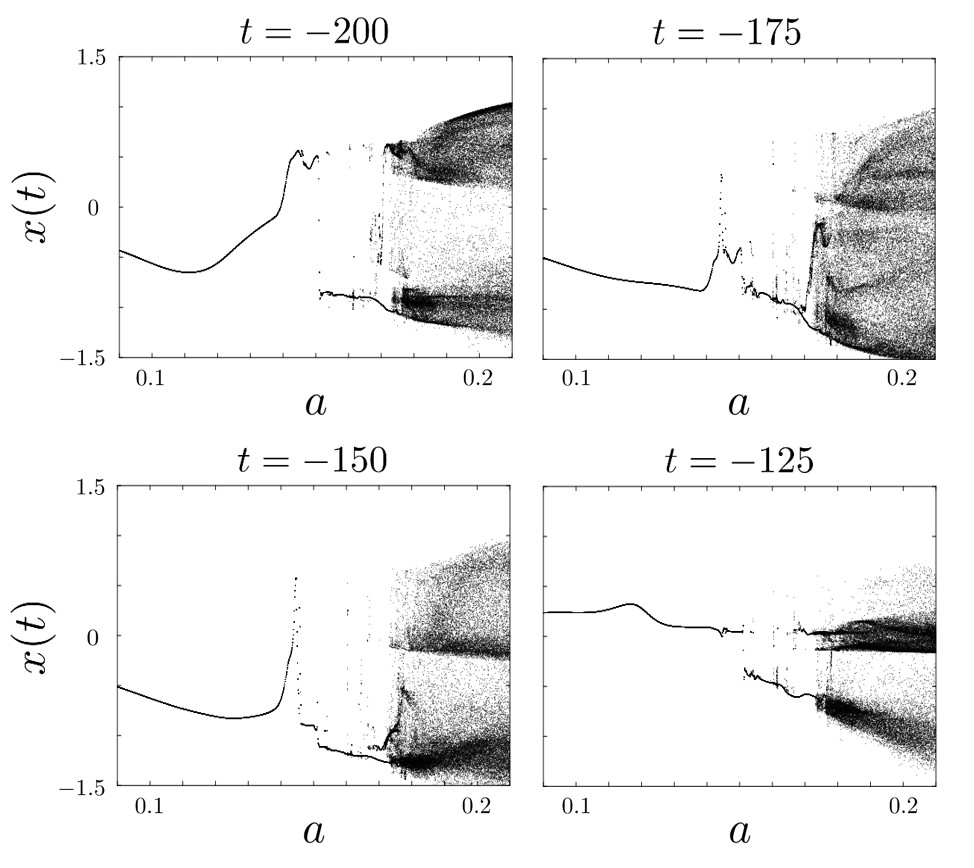}
    \caption{Snapshots of bifurcation diagrams of random pullback attractors for $\sigma=0.15$ as the parameter $a$ is varied under a fixed noise realization. For all parameters, random pullback attractors are computed using a fixed pullback time $t_p = 4\times10^4$. The numerical scheme follows that used in Fig. \ref{fig:Bifur}.}
    \label{fig:supp_bifur_pullback}
\end{figure}

We construct a bifurcation diagram of random pullback attractors for the SMG. Plotting trajectories starting from many initial conditions under a fixed noise realization for each value of the parameter $a$, the bifurcation of random pullback attractors is successfully visualized. Similar to the attractors themselves, the bifurcation diagram evolves in time. In Fig. \ref{fig:supp_bifur_pullback}, the pseudo-periodic oscillation of the distribution $\rho_t$ of all possible trajectories appears near $a = 0.14$ and $\sigma=0.15$, illustrating the onset of SR behavior. 

\section{Linear mode analysis of unstable spirals}
\begin{figure}[htbp]
    \centering
    \includegraphics[scale=0.35]{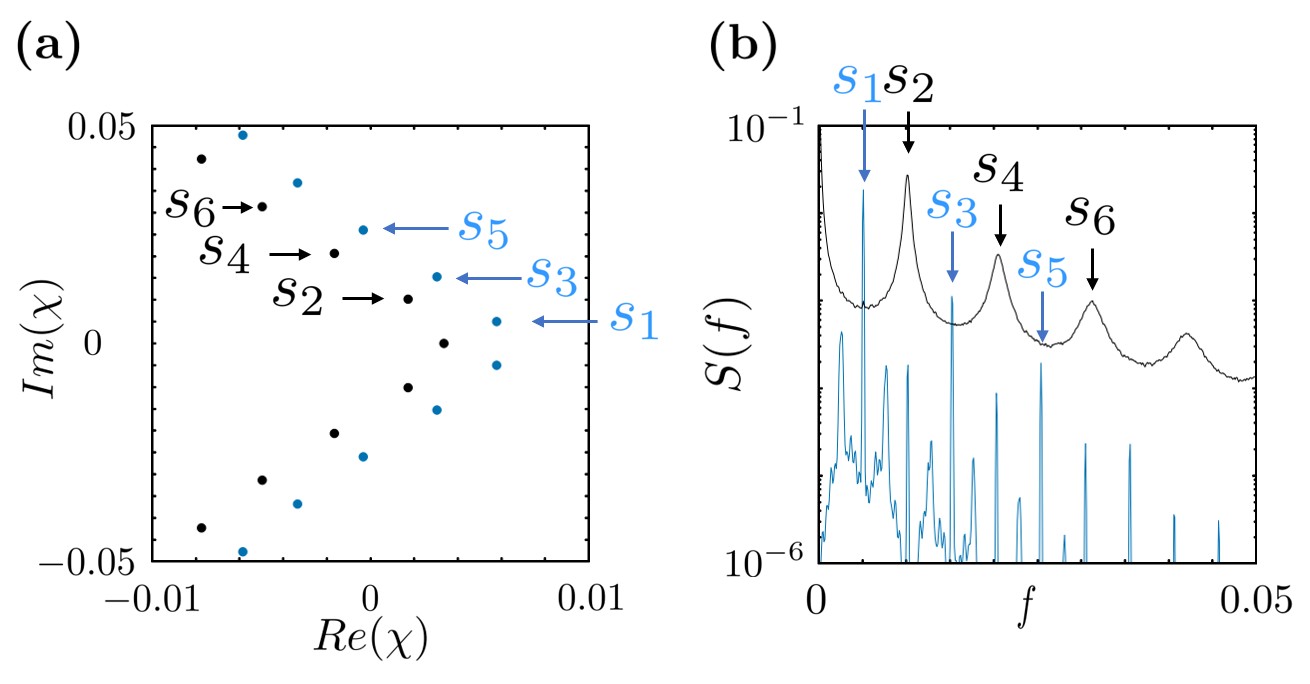}
    \caption{(a) Roots of linear modes $\chi = \lambda_n$ (black) and $ chi = \mu_n$ (blue) at $a = a_1 = 0.14$. (b) Power spectra for $a=a_1=0.14$ with $\sigma=0$ (blue) and optimal noise $\sigma = \sigma^*$ (black). }
    \label{fig:supp_linearmode}
\end{figure}

When the delay $\tau$ is sufficiently large, the peaks in the power spectra of the MG correspond to the linear modes around fixed points \cite{Mensour_Longtin}. 
The characteristic equations at fixed points $x=0,\pm x^*$, where $x^*=(\frac{a-b}{b})^{1/c}$,
\begin{align}
\begin{array}{ll}
\chi +b-a e^{-\chi \tau}=0 & (x=0),\\
\chi +b-(\frac{cb^2}{a}-(c-1)b) e^{-\chi \tau}=0 & (x=\pm x^*),\\
\end{array}
\end{align}
have an infinite number of roots $\chi=\lambda_n,\mu_n ~(n=0,\pm1 \pm 2,\ldots)$ given by 
\begin{align}
\begin{array}{ll}
\lambda_n = -b + \frac{1}{\tau} W_n\left[a\tau\exp(b\tau)\right] &(x=0), \\
\mu_n = -b + \frac{1}{\tau}W_n\left[(\frac{cb^2}{a}-(c-1)b)\tau\exp(b\tau)\right] &(x=\pm x^*), 
\end{array}
\end{align}
where $W_n(z)$ denotes the Lambert W-function \cite{corless1996lambert}. Using the asymptotic expansion of the Lambert W-function \cite{Amann2007}, the frequencies associated with the linear modes are given by the imaginary parts of $\lambda_n$ and $\mu_n$:
\begin{equation}
\begin{array}{ll}
    s_{2n} &:= \frac{1}{2\pi} Im(\lambda_n) \\
    &\simeq  \frac{2n}{2\tau} \left[ 1-\frac{1}{b \tau + \ln(a\tau)} \right] ~~~~~~~~~~~~~~~~~~~~~~~~ (x=0), \\
    s_{2n+1} &:= \frac{1}{2\pi} Im(\mu_n) \\
    &\simeq \frac{2n+1}{2\tau} \left[ 1-\frac{1}{b \tau + \ln(|\frac{cb^2}{a}-(c-1)b|\tau)} \right] ~~~~~ (x=\pm x^*).
    \end{array}
\end{equation}
The even modes $s_{2n}$ describe the unstable spiral around the origin, while the odd modes $s_{2n+1}$ characterize spirals around $x=\pm x^*$.  The deterministic dynamics dominated by odd modes. When the noise is added, the odd modes are suppressed and the even modes become dominant (Fig. \ref{fig:supp_linearmode}), producing SR with resonant frequencies associated with the even linear modes. 

\section{Universality of chaotic stochastic resonance}
\label{sec:Universality}

Chaotic SR arises not only in time-delayed systems exhibiting coherence resonance, but also in periodically forced systems exhibiting classical SR and in excitable systems exhibiting coherence resonance. 

\begin{figure}[htbp]
    \centering
    \includegraphics[width=0.92\linewidth]{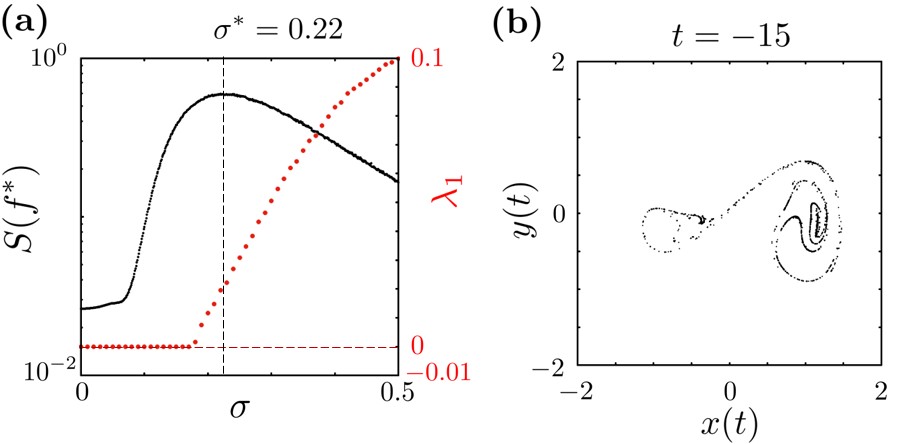}
    \caption{
    Chaotic SR in the Duffing equation with $\gamma=0.2$, $\varepsilon=0.3$, and $\omega=0.1$. (a) Power at the resonant frequency $S(f^*)$, $f^*\simeq 0.0159$ (black points), and the largest random LE $\lambda_1$ (red points) as functions of $\sigma$. At the resonance point $\sigma=\sigma^* \simeq 0.22$, the largest random LE $\lambda_1 \simeq 0.018>0$. (b) Snapshot of the random pullback attractor at the resonance point, computed from $2\times10^4$ initial conditions uniformly distributed over  $[-2,2]\times[-2,2]$ with a pullback time $t_p = 400$. }
    \label{fig:Duffing}
\end{figure}

To illustrate  classical SR, we consider an overdamped dynamics in a double-well potential,
\begin{equation}
dx=(x - x^3 )ds.
\label{eqs:odd}
\end{equation} 
Upon adding a slow periodic external force and Gaussian noise, we obtain
\begin{equation}
dx=\left[x - x^3 + \varepsilon \cos(\Omega s)\right]ds+\sigma' dW_s,
\label{eqs:odds}
\end{equation}
where $\sigma'$ represents the noise intensity, and $W_s$ denotes the Wiener process. The external force is specified by  the amplitude $\varepsilon$ and the angular frequency $\Omega$. For $\varepsilon=0.1$ and $\Omega=2\pi/100$, we observe stable SR in Eq. (\ref{eqs:odds}) at $\sigma'\simeq 0.354$  \cite{SR_review}.
Next, we consider the underdamped dynamics 
\begin{eqnarray}
    dx &=& ydt\nonumber\\
    dy &=& ( -\gamma y + x - x^3 ) dt, 
    \label{eqs:udd}
\end{eqnarray}
where $\gamma$ is the damping rate. In the overdamped limit ($\gamma\rightarrow\infty$,  $t=\gamma s$, and $dy/dt\rightarrow 0$), Eq. (\ref{eqs:udd}) reduces to Eq. (\ref{eqs:odd}). By  adding a slow periodic external force and Gaussian noise, we obtain the stochastic Duffing equation given by 
\begin{eqnarray}
    dx &=& ydt\nonumber\\
    dy &=& \left[ -\gamma y + x - x^3 + \varepsilon \cos(\omega t) \right]dt +\sigma dW_t,
    \label{eqs:udds}
\end{eqnarray}
where $\sigma$ represents the noise intensity, and $W_t$ denotes the Wiener process. The external force is specified by the amplitude $\varepsilon$ and the angular frequency $\omega$. For $\gamma=0.2$, $\varepsilon=0.3$, and $\omega=0.1$, we observe chaotic SR in Eq. (\ref{eqs:udds}) at $\sigma\simeq 0.22$ (see FIG.\ref{fig:Duffing}). Similar phenomena have been reported in both numerical and laboratory experiments  \cite{wiebe2014co}.

\begin{figure}[htbp]
    \centering
    \includegraphics[width=0.92\linewidth]{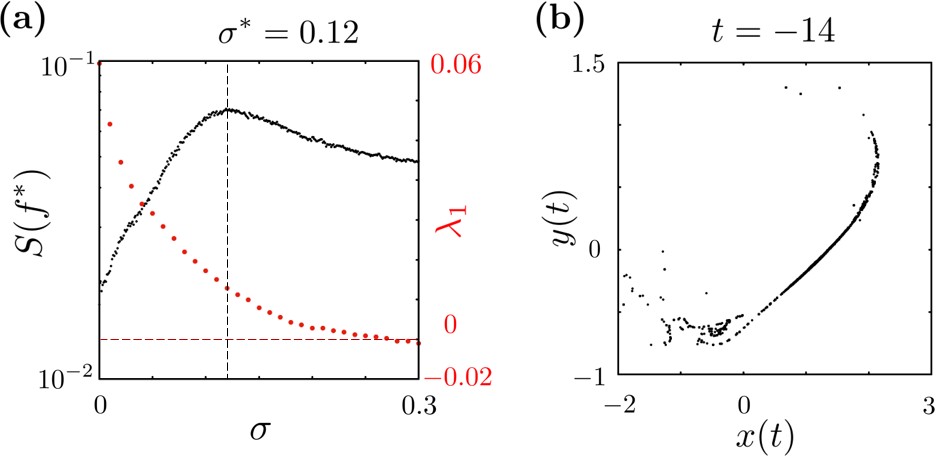}
    \caption{
    Chaotic SR in the underdamped FitzHugh--Nagumo equation with $\gamma=0.41$, $\varepsilon=0.1$, and $a=1.1$. (a) Power at the resonant frequency $S(f^*)$, $f^*\simeq 0.037$  (black points), and the largest LE $\lambda_1$ (red points) as functions of $\sigma$. At the resonance point $\sigma=\sigma^* \simeq 0.12$, the Lyapunov exponent is $\lambda\simeq 0.003>0$.  (b) Snapshot of the random pullback attractor, projected onto $(x(t), y(t))$ plane, at the resonance point $\sigma = \sigma^*$, computed from $2\times10^4$ initial conditions uniformly distributed over $[-2,2]\times[-2,2] \times [-2,2]$ with a pullback time $t_p = 400$.}
    \label{fig:FitzHugh}
\end{figure}

For coherence resonance, we consider the FitzHugh-Nagumo equation,
\begin{eqnarray}
\varepsilon dx &=& (x - \frac{x^3}{3} -y) ds, \nonumber\\
dy &=& (x+a) ds
\label{eqs:odfn}
\end{eqnarray}
where $\varepsilon$ represents the timescale separation  between $x$ and $y$ and $a$ is the excitation  threshold.  With additive Gaussian noise, we obtain  
\begin{eqnarray}
\varepsilon dx &=& (x - \frac{x^3}{3} -y) ds, \nonumber\\
dy &=& (x+a) ds +\sigma'dW_s
\label{eqs:odfns}
\end{eqnarray}
where $\sigma'$ represents the noise intensity, and $W_s$ denotes the Wiener process. For $\varepsilon=0.01$ and $a=1.05$, we observe stable SR in Eq. (\ref{eqs:odfns}) at $\sigma'\simeq 0.07$ \cite{Pikovsky_Kurths}.
Next we consider the underdamped FitzHugh-Nagumo  equation \cite{scott1975electrophysics}
\begin{eqnarray}
dx &=& z dt, \nonumber\\
dy &=& \frac{\varepsilon}{\gamma} (x+a) dt \nonumber\\
dz &=& (x-\frac{x^3}{3}-y-\gamma z)dt, 
\label{eqs:udfn}
\end{eqnarray}
where $\gamma$ is the damping rate, and the overdamped limit ($\gamma\rightarrow\infty$, $t=\gamma s/\varepsilon$, and $dz/dt\rightarrow 0$) reduces to  (\ref{eqs:odfn}). 
With additive Gaussian noise, we obtain 
\begin{eqnarray}
dx &=& z dt, \nonumber\\
dy &=& \frac{\varepsilon}{\gamma}(x+a) dt +\sigma dW_t\nonumber\\
dz &=& (x-\frac{x^3}{3}-y-\gamma z)dt, 
\label{eqs:udfns}
\end{eqnarray}
where $\sigma$ represents the noise intensity, and $W_t$ denotes the Wiener process. For $\gamma=0.41$, $\varepsilon=0.1$, and $a=1.1$, we observe chaotic SR in Eq. (\ref{eqs:udfns}) at $\sigma\simeq 0.12$ (see Fig. \ref{fig:FitzHugh}).

\bibliography{thesis_ref}

\end{document}